%% file: eulerSieve.tex
\documentclass[10pt]{article}

\usepackage{booktabs} 

\usepackage[ruled]{algorithm2e} 

\SetAlFnt{\small}
\SetAlCapFnt{\small}
\SetAlCapNameFnt{\small}
\SetAlCapHSkip{0pt}
\IncMargin{-\parindent}

\usepackage{url}
\usepackage{amsmath,latexsym,amssymb}
\usepackage{amsfonts}

\usepackage{xcolor}
\usepackage{tcolorbox}
\usepackage{geometry}
\usepackage[inline]{enumitem}
\usepackage{tabu}

\newcommand{\code}[1]{\texttt{\small #1}}

\newcommand{\primes}{\mathcal{P}}

\title
{Three Euler's Sieves and a Fast Prime Generator \\ (Functional Pearl)
\vspace{-3mm}}

\author{Ivano Salvo$^{\dag}$, Agnese~Pacifico$^\ddag$ \medskip
\\
$^\dag$ Departement of Computer Science, Sapienza University of Rome\\
$^\ddag$ Departement of Mathematics, Sapienza University of Rome\medskip
\vspace{-3mm}
}
\date{}

\begin{document}
\maketitle

%

\begin{abstract}
The Euler's Sieve refines the Sieve of Eratosthenes to compute prime numbers, 
by crossing off each non prime number just once. Euler's Sieve is considered 
hard to be faithfully and efficiently coded as a purely functional stream based program. 
We propose three Haskell programs implementing the Euler's Sieve, 
all based on the idea of generating just once each composite to be crossed off. 
Their faithfulness with respect to the Euler's Sieve is 
up to costly stream unions 
imposed by the sequential nature of streams.
Our programs outperform classical na\"ive stream based prime generators such as trial division, but they 
are asymptotically worse than the O'Neill `faithful' Sieve of Eratosthenes.   
To circumvent the bottleneck of union of streams, 
we integrate our techniques inside the O'Neill program, thus obtaining a fast prime generator based on the Euler's Sieve and priority queues. 
%
\end{abstract}
\maketitle


\input{intro}

\input{eratostheneBird}

\input{hammingRevisited}

\input{eulerGenuineSieve}

\input{classicalEuler}

\input{pitStop}
\input{wheelONeal}
\input{expres}

\input{conclu}


\bibliographystyle{alpha}
\bibliography{biblio}

\end{document}

%% file: intro.tex
\section{Introduction}
\label{sec:intro}

The generation of the stream of prime numbers is a classical and well-studied problem, that has been deeply 
investigated for a long time (see for example~\cite{pritchard87}). 
The first algorithm, and probably the best kwown, is the Sieve of Eratosthenes: 
in this algorithm,
after discovering a new prime $p$, all multiples  of $p$, $\{p\cdot k~|~k\in\mathbb{N}_{\geq 2}\}$, 
are crossed off as non-primes. 
Since composites less than $p^2$ have at least a prime factor $p'<p$, 
this process can be quicken, by starting from $p^2$, because smaller composites have been already crossed off as multiples of some 
$p'<p$. The complexity of this algorithm is ${\cal O}(n \ln \ln n)$ to generate all primes less than $n$.  

Nevertheless, a lot of useless work is carried out in the execution of the Sieve of Eratosthenes: as an example, 
crossing off multiples of 3, we cross off again even numbers such as 12, 18, 24, \ldots that have 
already  been crossed off as multiples of 2. This suggests a refinement of this algorithm, the Euler's Sieve, 
in which during the $k^{th}$ iteration, only numbers that are multiple of $p_k$, 
but not multiple of $p_1,\ldots,p_{k-1}$ are crossed off. 
The asymptotic speed-up of the Euler's Sieve with respect to the Sieve of Eratosthenes is $\ln \ln n$,  
that is (on average) the number of distinct prime factors of $n$. Therefore its complexity is ${\cal O}(n)$.

The natural implementation of the Sieve of Eratosthenes requires direct access to prime candidates to be crossed off, 
and hence array based imperative programs reflects the original procedure 
more than stream based purely functional programs. 
As a matter of fact, despite several elegant and concise Haskell programs that lazily generates the stream of primes, 
a `faithful' (from the point of view of both efficiency and the computations actually performed)
functional implementation of the Sieve of Eratosthenes is far from trivial~\cite{sieve09}.

The so-called stream-based Sieve of Eratosthenes (see Fig.~\ref{fig:unfaithful-sieve}) is still on the official home page of Haskell 
\cite{HHP} 
as a paradigmatic example of the conciseness and the level of abstraction 
that one can achieve in Haskell. 
Since its first appearance \cite{turner75}, however, its performances appeared not to be worthy of its cleanliness, 
as they are rather poor compared to other simple algorithms that lazily generate the stream of primes, such as trial division
(e.g., see \cite{haskellPrimes}). 

\begin{figure}[h]
\begin{center}
\begin{minipage}{0.8\linewidth}
\begin{tcolorbox}[colback=white, boxrule=0pt, toprule=1pt, bottomrule=1pt, left=0pt]
{\small
\begin{verbatim}
   primes = filterPrime [2..] where 
      filterPrime (p:xs) = 
            p : filterPrime [x | x <- xs, x `mod` p /= 0]
\end{verbatim}
}
\end{tcolorbox}
\end{minipage}
\end{center}
\vspace{-6mm}
\caption{The `unfaithful' Sieve of Eratosthenes \cite{turner75}.}
\label{fig:unfaithful-sieve}
\vspace{-1mm}
\end{figure}

Even in imperative languages, to make the Euler's Sieve efficient requires some ingenuity and it is less obvious than 
for the Sieve of Eratosthenes (e.g., see~\cite{js90}). 
The traditional Haskell program implementing the Euler's Sieve 
(see Fig.~\ref{fig:classical-sieve-of-euler})
performs even worse than the unfaithful Sieve of Eratosthenes. 
The problem is that the prime $p_k$, before being recognised as prime, 
as the head of the not yet crossed off numbers, must survive to $k-1$ stream `complementation' (via the \code{minus} function). 
In turn, each stream complementation generates a new stream and this generation is a never-ending process, 
causing very soon also a memory explosion problem.

\begin{figure}[h]
\begin{center}
\begin{minipage}{0.8\linewidth}
\begin{tcolorbox}[colback=white, boxrule=0pt, toprule=1pt, bottomrule=1pt, left=0pt]
{\small
\begin{verbatim}
   minus xs@(x:txs) ys@(y:tys) 
      |   x < y   = x : minus txs ys
      |   x > y   = minus xs txs
      | otherwise = minus txs tys  

   primes = eulerSieve [2..] where
      eulerSieve cs@(p:tcs) = p:eulerSieve (tcs `minus` map (p*) cs)
\end{verbatim}
}
\end{tcolorbox}
\end{minipage}
\end{center}
\vspace{-6mm}
\caption{The classical Euler's Sieve in Haskell \cite{haskellPrimes}.}
\label{fig:classical-sieve-of-euler}
\vspace{-1mm}
\end{figure}

The Euler's Sieve is  considered hard to be coded in a stream based fashion or even impossible in principle 
(e.g., see \cite{haskellPrimes}, Section 5.3) because, differently from multiples, numbers to be crossed off 
depend on all $p_1, \ldots p_k$ and they appear not to be efficiently computable from the stream of primes under construction. 

\subsection{The Euler's Sieve, Formally}
\label{sec:eulerFormal}
Euler's Sieve can be formally defined by specifying the set of numbers to be crossed off and the set of those 
that survive as prime candidates at each iteration of the algorithm. 
We start by giving some notation and definitions. 

Let $A\cdot B=\{a\cdot b~|~ a\in A, b\in B\}$ 
be the set of all products of numbers in $A$ and $B$. We write $a\cdot B$ for $\{a\}\cdot B$.  
We use the notation $p\mid n$ (resp. $p\nmid n$) to mean that $p$ is (resp. is not) a factor of $n$.
We denote with $\primes$ the sequence $[2, 3, 5, 7, \ldots, p_k, \ldots]$ of prime numbers that, 
when convenient, we regard as a set. We denote with $C(P)$ the set of composites of a set of primes $P\subseteq\primes$.
Given a sequence $A$, we will denote with $A|_k$ the prefix of its first $k$ elements 
and with $A|^k$ the suffix starting at its $k$ element.
Accordingly, we stipulate that $p_1$ is $2$ and we denote with $\primes|_k$ the first $k$ primes, and with 
$\primes|^k$ the suffix of primes starting in $p_k$.

For a natural number $k>0$, $E_k$ is the set of natural numbers crossed off at the $k^{th}$ iteration of the 
Euler's Sieve (`$E$' stands for `erased'), that is $E_k=\{n\in\mathbb{N}_{\geq 2}~|~\forall p\in \primes|_{k-1} p\nmid n~ ~\& ~p_k\mid n\}$. For $k\geq 0$, $S_k$ is the set of naturals that are still candidate to be prime after the 
$k^{th}$ iteration of the Euler's Sieve (`$S$' stands for `survived'), that is $S_k=\{n\in\mathbb{N}_{\geq 2}~|~\forall p\in \primes|_{k}~p\nmid n\}$.
In view of our recursive programs, it is useful to give an inductive definition of these sequences of sets, as follows:
\begin{equation}
E_0=\varnothing ~~~~~ S_0 = \mathbb{N}_{\geq 2} ~~~~~ 
E_{k+1} = p_{k+1}\cdot S_{k}|^{k+1}   ~~~~~ S_{k+1} = S_{k}\setminus E_{k+1}  
\label{eq:eksk}
\end{equation}

Clearly, we have that $S_k\supset S_{k'}$ for $k<k'$, and $E_k\cap E_{k'}=\varnothing$ for $k\not=k'$. 
Observe also that the property $S_k \cup \bigcup_{i=0}^k E_i = \mathbb{N}_{\geq 2}$ is invariant for all $k$. 
Essentially, the Euler's Sieve in Fig.~\ref{fig:classical-sieve-of-euler} computes the stream of prime numbers 
accordingly to the fact that  $\primes=\bigcap_{k=0}^{\infty} S_k$, 
using our mutual inductive definition of the sequences of sets $E_k$ and $S_k$. 
The set of all composite numbers, $C(\primes)$ can be characterised as $\bigcup_{k=0}^\infty E_k$. 

\subsection{Our Contributions}
Inspired mainly by a stream based Sieve of Eratosthenes attributed to Richard~Bird in \cite{sieve09} 
(see Fig.~\ref{fig:sieveBird}), 
we present three Haskell programs implementing the Euler's Sieve. All these programs  
inherit from the Bird program a change of perspective: we look at primes as the ground of composites and therefore 
our goal becomes to generate each composite just once. 

Our first solution is based on a new solution to a generalisation of the Hamming problem~\cite{birdWadler88}. 
Our second solution makes use of wheels as a tool to generate composites to be crossed off, rather than 
to generate primes, as they have  been usually used in prime generators~\cite{runciman97}. 
Our third solution essentially applies Equation (\ref{eq:eksk}), making efficient the idea behind the Euler's Sieve of Fig.~\ref{fig:classical-sieve-of-euler}. 

The faithfulness of these programs with respect to the original Euler's Sieve is up 
to the bottleneck of costly union operations 
required because of the sequential nature of streams.
To overcome such problem, we have finally integrated our second and third solutions   
into the priority-queue based `faithful' Sieve of Eratosthenes in \cite{sieve09}. 
In particular, we regard our priority-queue wheel based Euler's Sieve as our fast prime generator, 
as it is fast enough and resistant to performance degradation due to memory (de)allocation 
when it computes milions of primes. 


%% file: eratostheneBird.tex
\section{Figure and Ground: Primes vs Composites}
\label{sec:seh}

As in Escher litographs and woodcuts, prime and composite numbers form a figure-ground picture~\cite{geb79}. 
In the Epilogue of the Melissa O'Neill paper \cite{sieve09} about faithful sieves, 
it is reported a brilliant purely stream based program attributed to Richard Bird (see Fig.~\ref{fig:sieveBird})  
essentially based on the following recursive equation:
\begin{equation}
\primes = \mathbb{N}_{\geq 2} \setminus C(\primes)
\label{eq:primes-composites}
\end{equation}
To make the computation productive, we extract the first prime, 2, and we characterise $C(\primes)$ as 
union of multiples of primes according to the Sieve of Eratosthenes, thus obtaining: 
\begin{equation}
\primes = \{2\}\cup\mathbb{N}_{\geq 3} \setminus \bigcup_{k=1}^{\infty} p_k\cdot\mathbb{N}_{\geq p_k}
\label{eq:primes-composites-2}
\end{equation}

Roughly speaking, Bird's sieve  simply computes the list of all composite numbers as 
the union of multiples of all numbers in the list of primes  
that, in turn, is under construction as the complement of composites numbers (`union' here means merge of ordered lists, 
possibly avoiding duplicates). As in the classical Sieve of Eratosthenes, 
in this program each composite number is crossed off once for each of its distinct prime factors. 
To be precise, here `crossed off' means {\em generated} in the list of composites. 
For example, 120 will be generated 3 times as a multiple of 2, 3, and 5. 


\begin{figure}[h]
\begin{center}
\begin{minipage}{0.8\linewidth}
\begin{tcolorbox}[colback=white, boxrule=0pt, toprule=1pt, bottomrule=1pt, left=0pt]
{\small
\begin{verbatim}
    union xs@(x:txs) ys@(y:tys)
       | x == y = x:union txs tys
       | x  < y = x:union txs ys
       | x  > y = y:union xs  tys 

    primes = 2:([3..] `minus` composites) where
        composites = foldr unionP [] [multiples p | p <- primes] 
        multiples n = map (n*) [n..]
        unionP (x:xs) ys = x:union xs ys
\end{verbatim}
}
\end{tcolorbox}
\end{minipage}
\end{center}
\vspace{-6mm}
\caption{Stream based Sieve of Eratosthenes by R.~Bird \cite{sieve09}}
\label{fig:sieveBird}
\vspace{-1mm}
\end{figure}

We observe that Bird's program uses a smart trick to make productive the computation of the union of a stream of streams, 
by using a union function  (we use the name \code{unionP}, where `\code{P}' stands for `productive') 
that always chooses the first element of the first stream and then 
proceeds as an usual \code{union}. 
In this particular case, we know that this is correct, because primes are a ordered list and 
each list of multiples of $p$ starts at $p^2$ and hence for $i<j$, we have $p_i^2<p_j^2$. 

Motivated by improving Bird's sieve by generating each composite number {\em just once}, in this functional pearl, 
we essentially look for a purely stream-based Haskell program implementing  the Euler's Sieve.

%% file: hammingRevisited.tex
\section{Primes as Background of Generalised Hamming Numbers}
\label{sec:hpr}
The problem of generating composites is tightly related to a generalised Hamming problem. 
Given a set of generators $P$ (usually, but not necessarily, primes), the problem consists of generating in increasing order 
the smallest set $H(P)$ such that $1\in H(P)$ and for all $p\in P$ and $h\in H(P)$, $p\cdot h\in H(P)$. 
Using our notations, $H(P)$ is the smallest set containing 1 and satisfying the equation $P\cdot H(P)=H(P)$.

In \cite{birdWadler88}, it is presented a solution to the classic version of the Hamming problem with $P=\{2,3,5\}$ 
that can be easily generalised to generate Hamming numbers starting from any list of generators 
of arbitrary length 
(Exercise 7.6.5 in \cite{birdWadler88}). Some solutions can be found in \cite{sp18}.

Unfortunately, such classical solutions do not serve to our main purpose, that is generate each  
composite number just once, because, for example 
computing $H(\{2,3,5\})$ they generate the number 30 six times (all permutations of factors of 30).

%
%

\subsection{The Hamming Problem Revisited}
\label{sec:cshp}
In the Afterwords of \cite{sp18}, we presented a solution to the Hamming problem that 
generates once each composite. 
%
%
We can do much better, however, thinking the set of $H(P)$ 
 as the smallest set satisfying the following equation ($p\in P$): 
\begin{equation} 
H(P) = p\cdot H(P) \cup H(P\setminus \{p\})
\vspace{-1mm}
\label{eq:hamming2}
\end{equation}

Observe that, if we are interested in $H'(P)=H(P)\setminus \{1\}$, 
from Equation (\ref{eq:hamming2}) we have $H'(P) = \{p\}\cup p\cdot H'(P) \cup H'(P\setminus \{p\})$. 
Moreover, if numbers in $P$ are primes, then $p\cdot H'(P) \cap H'(P\setminus \{p\})=\varnothing$, because 
all numbers in $p \cdot H'(P)$ have $p$ as a prime factor, whereas 
numbers in $H'(P\setminus \{p\})$ do not. This implies that we generate just once each composite as desired.
This equation leads to a small (and highly circular) Haskell program, 
that always outperforms the classical solution to the Hamming problem (to be precise, function \code{hamming} 
in Fig.~\ref{fig:composti-circular} never generates 1). 

\begin{figure}[h]
\begin{center}
\begin{minipage}{0.68\linewidth}
\begin{tcolorbox}[colback=white, boxrule=0pt, toprule=1pt, bottomrule=1pt, left=0pt]
{\small
\begin{verbatim}
   dUnion xs@(x:txs) ys@(y:tys)  
      | x <  y    = x:dUnion txs ys
      | otherwise = y:dUnion xs tys
   dUnion xs [] = xs

   hamming [] = []
   hamming (x:xs) = hmngs where
      hmngs = x:map (x*) hmngs `dUnion` hamming xs
\end{verbatim}
}
\end{tcolorbox}
\end{minipage}
\end{center}
\vspace{-5mm}
\caption{A solution to the Hamming problem that generates each number exactly once.}
\label{fig:composti-circular}
\end{figure}

Thanks to the fact that we generate disjoint streams of composites, 
we can also slightly optimise the \code{union} function: the function \code{dUnion} (`\code{d}' stands 
for `disjoint'), assuming
as precondition that its parameters are disjoint ordered lists, avoids to consider the case \code{x==y} 
that never occurs: 
this small optimisation has a significant impact on running time of the function \code{hamming}.

\subsection{Euler's Sieve from Hamming Numbers}
\label{sec:cshp}

Function \code{hamming} correctly computes also Hamming numbers of an infinite list of generators, such as the stream of primes.
Moreover, its recursive calls are tightly related to the iterations of the Euler's Sieve.
As a matter of fact, $H(\primes)=\mathbb{N}$ and $2\cdot H(\primes)$ corresponds to all even numbers, that are in turn 
$\{2\}\cup E_1$, that is numbers crossed off in the first iteration 
of the Euler's Sieve (in this particular case, this also corresponds to the set of numbers crossed off 
by the first iteration of the Sieve of Eratosthenes). 
Similarly, the set $3\cdot H(\primes|^1\})$ is $\{3\}\cup E_2$, that is all multiples of 3 that are not 
multiples of 2, again 3 plus the set of numbers crossed off in the second iteration of the Euler's Sieve, and so on. 

Stemming from this solution of the Hamming problem, 
we can write 
an efficient lazy generator of prime numbers that ideally implements the Euler's Sieve in Haskell (see Fig.~\ref{fig:primes-hammings}). 
Since we are interested in computing the set of composites $C(\primes)=H(\primes)\setminus \primes$, 
we need just to avoid to insert generators (i.e. prime numbers) in the resulting list of composites. 
This makes the code just a bit more involved, because the suffix $\primes|^{k+1}$ of prime numbers 
is needed to compute $C (\primes|^k)$, but they are not present in the list of composites $C(\primes|^{k+1})$
that we get from the recursive call. 
Therefore, they must be reinserted before computing $C (\primes|^k)$.

\begin{figure}[h]
\begin{center}
\begin{minipage}{0.68\linewidth}
\begin{tcolorbox}[colback=white, boxrule=0pt, toprule=1pt, bottomrule=1pt, left=0pt]
{\small
\begin{verbatim}
     sMinus xs@(x:txs) ys@(y:tys)
        | x==y       = sMinus txs tys
        | otherwise  = x:sMinus txs ys
      
     composites (x:xs) = cmpsts where
         cmpsts = (x*x):map (x*) (xs `dUnion` cmpsts) 
                    `dUnion` composites xs

     primes = 2:([3..] `sMinus` composites primes)
\end{verbatim}
}
\end{tcolorbox}
\end{minipage}
\end{center}
\vspace{-5mm}
\caption{{\sf H}: the Euler's Sieve based on Hamming Numbers (of Primes)}
\label{fig:primes-hammings}
\end{figure}

We observe that in Fig.~\ref{fig:primes-hammings}
we use the function \code{sMinus} (instead of the standard \code{minus}, `\code{s}' stands for `subset').  
Function \code{sMinus} assumes as a precondition that input lists are ordered and the set of elements of the second list is 
contained in the first one: under this precondition, we can avoid to check the case \code{x>y}, that never happens. 
Similarly to \code{dUnion}, this small optimisation has a significant impact on the running time of this program.

Unfortunately, each composite number is generated just once, but several comparisons in nested calls of the \code{dUnion} function 
are needed for a number before joining the list of composites. This is, of course, the main reason why this program 
cannot achieve the expected speed-up. 
%
%

%% file: eulerGenuineSieve.tex




\begin{figure}[b]
\begin{center}
\begin{minipage}{0.80\linewidth}
\begin{tcolorbox}[colback=white, boxrule=0pt, toprule=1pt, bottomrule=1pt, left=0pt]
{\small
\begin{verbatim}
   primes = 2:([3..] `sMinus` composites) where
       composites = foldr dUnionP [] [wP p | p <- primes] 
       wP p = map (p*) (spin (wheel (takeWhile (<p) primes) p) p)
\end{verbatim}
}
\end{tcolorbox}
\end{minipage}
\end{center}
\vspace{-5mm}
\caption{The na\"ive Euler's Sieve using wheels.}
\label{fig:wheel-Sieve}
\vspace{-4mm}
\end{figure}
 
\section{Reinventing Wheels}
\label{sec:wheels}

Wheels are a typical tool to generate primes (e.g., see~\cite{pritchard82}). 
In this paper, we consider the sequence of wheels $w_0, w_1, \ldots$ such that 
by `rolling' the wheel $w_k$ starting from $p_{k+1}$, we efficiently generate all numbers 
that are not multiples of $p_1, \ldots, p_k$, that are in turn $S_k|^{k+1}$.

We define $w_0$ as the sequence $[1]$, that starting in 2 generates $\mathbb{N}_{\geq 2}=S_0$.   
Let $\Pi_k$ be the product $p_1 p_2\ldots p_k$ of the first $k$ primes. 
Let $[q_1,\ldots, q_m]$ be the ordered sequence of numbers in the interval 
$[p_{k+1}\ ..\ p_{k+1}+\Pi_k]$ such that 
$p_i \nmid q_j$ for all $i\in [1\ ..\ k]$ and $j\in [1\ .. \ m]$. The wheel $w_{k}$  
of circumference $\Pi_k$ starting in $p_{k+1}$ is the sequence $[q_1-p_{k+1}, q_2-q_1, \ldots, \Pi_k-q_m]$. 

 
Wheels of arbitrary size are used in \cite{runciman97} to generate all prime numbers following 
the Wheel Sieve in \cite{pritchard82}. 
Unfortunately, the resulting Haskell programs are not so performant.
More usually, a fixed size pre-computed wheel is used to dramatically improve the running time of 
a prime generator, even without changing its asymptotic complexity (see Section~\ref{sec:pitStop}).

Again, by changing our point of view, we can use wheels to generate composites to be crossed off, 
rather than primes as in \cite{runciman97}. 
Since rolling the wheel $w_{k}$ starting from $p_{k+1}$, we get the sequence $S_k|^{k+1}$, 
we can use $w_k$ to generate composites to be crossed off after finding a new prime $p_{k+1}$ 
as $p_{k+1}\cdot S_k|^{k+1}=E_{k+1}$. 
Therefore, we can replace multiples of a prime $p_k$ 
in the Bird's Sieve of Fig.~\ref{fig:sieveBird} 
with $E_k$ just by rolling the wheel $w_{k-1}$ (see Fig.~\ref{fig:wheel-Sieve}). 

%


Even though this program performs quite well and it does not suffer from huge memory allocation 
of that one in Fig.~\ref{fig:primes-hammings}, we present it mainly because it shows very clearly the idea of using wheels 
to implement the Euler's Sieve. However, it contains a couple of evident inefficiencies: 
\begin{enumerate*}
\item function \code{wP} is called for each prime number $p$, and it recomputes at each invocation the prefix of primes up to $p$;
\item at each invocation, function \code{wheel} computes the wheel $w_k$ from scratch, taking as input primes $p_1, \ldots, p_k$ 
and therefore it has to perform a trial division on the finite interval of naturals $[p_{k+1}\ ..\ \Pi_k+p_{k+1}]$ (that becomes huge also for relatively small $k$).  
\end{enumerate*}

\begin{figure}[t]
\begin{center}
\begin{minipage}{0.72\linewidth}
\begin{tcolorbox}[colback=white, boxrule=0pt, toprule=1pt, bottomrule=1pt, left=0pt]
{\small
\begin{verbatim}
   nextWheel [] _ _      = [1]
   nextWheel (w:ws) p np = nWAux (rep p (ws++[w])) np p where
      nWAux    []  _ _  = []
      nWAux   [w]  _ _  = [w]
      nWAux (w:ws) s p  = 
        | mod (w+s) p == 0 = nWAux ((w+head ws):(tail ws)) s p
        | otherwise        = w:nWAux ws (w+s) p
      rep 0  _ = []
      rep n xs = xs ++ rep (n-1) xs

   nextWheel1 ws@(w:_) p = nextWheel ws p (p+w)  
   circ w = w ++ circ w
   spin (w:ws) n = n:spin ws (n+w)    
\end{verbatim}
}
\end{tcolorbox}
\end{minipage}
\end{center}
\vspace{-5mm}
\caption{Computing wheels, incrementally.}
\label{fig:wheels}
\vspace{-1mm}
\end{figure}

Both these two computations can be quicken by saving information on parameters of the function \code{composites}. 
In particular, 
since we need all wheels $w_1, w_2, \ldots, w_k, \ldots$, 
we add a wheel as a parameter in order to compute the wheel $w_{k+1}$ from the wheel $w_k$ as in \cite{runciman97}. 

The key observation is that the wheel $w_{k+1}$ consists of $p_{k+1}$ copies of $w_k$ and merging intervals when we hit a 
multiple of $p_{k+1}$. 
Instead of just picking the stream of wheels defined in \cite{runciman97}, 
in our programs based on explicit recursion, we find 
convenient to use function \code{nextWheel} as in Fig.~\ref{fig:wheels}.
Usually, we find convenient use the  
function \code{nextWheel1} that needs just one prime as parameter. Its correctness depends on the fact that having $p_{k}$ and 
$w_k$, $p_{k+1}$ is always $p_k+(head ~w_k)$. 
Finally, function \code{circ} in Fig.~\ref{fig:wheels} defines the repetitive stream $w^*$ generated by the wheel $w$, 
and function \code{spin} rolls a wheel (usually made repetitive by \code{circ}) starting from a given number.

\begin{figure}[h]
\begin{center}
\begin{minipage}{0.80\linewidth}
\begin{tcolorbox}[colback=white, boxrule=0pt, toprule=1pt, bottomrule=1pt, left=0pt]
{\small
\begin{verbatim}
   composites (p:ps) w = 
       map (p*) (spin (circ w) p) `dUnionP` composites ps w' where 
       w' = nextWheel1 w p
       dUnionP (x:xs) ys = x : dUnion xs ys
   
   primes = 2:([3..] `sMinus` (composites primes [1]))
\end{verbatim}
}
\end{tcolorbox}
\end{minipage}
\end{center}
\vspace{-5mm}
\caption{{\sf W}: The wheel based Euler's Sieve.}
\label{fig:wheel-Sieve-2}
\vspace{-1mm}
\end{figure}

As an example, taking as input the wheel $w_2=[2,4]$ that avoids to generate multiples of 2 and 3 starting from 5, 
and the primes 5 and 7, function \code{nextWheel w2 5 7} returns the wheel $w_3=[4,2,4,2,4,6,2,6]$ 
that avoids to generate multiples of 2, 3, and 5 starting from 7 as follows:
\begin{enumerate*}
\item 
first it makes 5 copies of the `shifted' wheel $w'_2=[4,2]$, that is $[4,2,4,2,4,2,4,2,4,2]$. Shifting is needed because we 
will roll this wheel starting from 7 and not from 5;
\item 
then it generates the corresponding sequence of numbers starting in 7, that is $[11,13,17,19,23,\mathbf{25},29,31,\mathbf{35},37]$;  and 
\item finally it `merges' delta's that corresponds to multiples of 5, thus obtaining $w_3=[4,2,4,2,4,2+4,2,4+2]$.
\end{enumerate*}
 
Having the wheel machinery and the already mentioned functions \code{sMinus} and \code{dUnion}, 
the result is the very small program {\sf W} of Fig.~\ref{fig:wheel-Sieve-2}, that again follows 
the figure-ground idea of prime-composite numbers
formalised in Equation (\ref{eq:primes-composites}) (\code{dUnionP} is for \code{dUnion} the analogous of \code{unionP} for \code{union}).

At each invocation of the function \code{composite (p:ps) w}, if \code{p} is the prime $p_k$ then \code{w} is the wheel $w_{k-1}$ 
and hence \code{map (p*) (spin (circ w) p)} is $p_k\cdot S_{k-1}|^k=E_k$.


%% file: classicalEuler.tex
\section{Back to the Origins} 
\label{sec:eulerOriginal}

Are wheels really necessary? Probably, the bad behaviour of the classical Haskell program  
implementing the Euler's Sieve in Fig.~\ref{fig:classical-sieve-of-euler} prevented us to start with 
a characterisation of composites to be crossed off along the lines of that program.

As we have seen in Section~\ref{sec:eulerFormal}, sets $E_k$ and $S_k$ can be defined by mutual induction, 
without the need of additional machinery such as wheels.
Again starting from Equation~(\ref{eq:primes-composites}), the idea is to compute the set of primes as the ground 
of composites as:
\begin{equation}
\primes = \mathbb{N}_{\geq 2}\setminus\bigcup_{k=1}^\infty E_k
\label{eq:es}
\end{equation}
By contrast, as already observed, the program of Fig.~\ref{fig:classical-sieve-of-euler} is essentially based on the equation  
$\primes = \bigcap_{k=0}^\infty S_k = \bigcap_{k=1}^\infty (\mathbb{N}_{\geq 2}\setminus E_k$)
that is set-theoretically equivalent, but it leads to a much more expensive computational process due to a deep nesting 
of stream complementations via the \code{minus} function.

As usual, the corresponding Haskell program {\sf ES} in Fig.~\ref{fig:efficient-euler-sieve} 
just rewrites the recursive Equations (\ref{eq:es}),  
extracting the first prime to make the lazy computation productive. 
At each invocation of \code{composites (p:ps) ss@(s:tss)}, if \code{p} is the prime $p_k$, \code{ss} is the iterator 
generating $S_{k-1}|^k$ and therefore \code{es = map (p*) ss} generates $E_{k}$, 
following the mutual induction schema defined in Equation (\ref{eq:eksk}).

\begin{figure}[h]
\begin{center}
\begin{minipage}{0.80\linewidth}
\begin{tcolorbox}[colback=white, boxrule=0pt, toprule=1pt, bottomrule=1pt, left=0pt]
{\small
\begin{verbatim}
   composites (p:ps) ss@(s:tss) = es `dUnionP` composites ps ss' where 
       es  = map (p*) ss
       ss' = tss `sMinus` es

   primes = 2:([3..] `sMinus` (composites primes [2..])) 
\end{verbatim}
}
\end{tcolorbox}
\end{minipage}
\end{center}
\vspace{-5mm}
\caption{{\sf ES}: The Euler's Sieve based on inductive computation of sets $E_k$ and $S_k$.}
\label{fig:efficient-euler-sieve}
\vspace{-1mm}
\end{figure}

%% file: pitStop.tex
\section{Pit Stop: Mounting Wheels on Sieves}
\label{sec:pitStop}

As noted in Section~\ref{sec:wheels}, a fixed size pre-computed wheel can improve the running time of 
a prime generator, even without changing its asymptotic complexity.
A common choice is the wheel $w_4$ that, starting from 11, generates all numbers that are not 
multiples of $2$, $3$, $5$, and $7$: this wheel avoids to check about 77\% of numbers for large $n$.    

This optimisation can be easily integrated in our programs, just modifying the definition of \code{primes}. 
For all programs, mounting the wheel $w_k$ prunes the set of candidate primes to be sieved, 
from $\mathbb{N}_{\geq 2}$ to $S_k|^{p_{k+1}}$. 
However, the impact on performance varies a lot among programs, due to the different meaning 
that mounting a pre-computed wheel has in the generation of composites.
 
In the Hamming Sieve of Fig.~\ref{fig:primes-hammings}, mounting the wheel $w_k$, requires 
to compute $C(\primes|^k)$ instead of $C(\primes)$: here, this is not only useful 
to significantly speed up its computation, but it is {\em necessary} for program correctness,  
to satisfy preconditions of function \code{sMinus}. This holds for all our programs

Mounting the wheel $w_k$ on the sieve {\sf W} of Fig.~\ref{fig:wheel-Sieve-2}  means just avoiding its first $k$ recursive 
calls, i.e. starting the computation from the wheel $w_k$ rather than from the wheel $w_0=[1]$. 
Since the computation of the first 4 small wheels is quite efficient, 
this optimisation has a limited impact on its performance.  

Mounting the wheel $w_k$ on the sieve {\sf ES} of Fig.~\ref{fig:efficient-euler-sieve} means starting the computation 
from $S_k|^{p_{k+1}}$: this is quite relevant, because in that program $S_4|^{11}$ is obtained 
by the first 4 stream complementations via \code{sMinus}, that are the most expensive.

In Fig.~\ref{fig:wheel-Hamming}, we give the new definitions of streams 
\code{primesH4} (for the Hamming based sieve {\sf H}), \code{primesW4} (for the wheel based sieve {\sf W}), and \code{primesES4} 
(for the sieve {\sf ES}). 

\begin{figure}[h]
\begin{center}
\begin{minipage}{0.80\linewidth}
\begin{tcolorbox}[colback=white, boxrule=0pt, toprule=1pt, bottomrule=1pt, left=0pt]
{\small
\begin{verbatim}
  s4  = spin (circ w4) 11
  ts4 = tail s4
  primesH4 = 2:3:5:7:11:ts4 `sMinus` composites (drop 4 primesH4)   
  primesW4 = 2:3:5:7:11:ts4 `sMinus` composites (drop 4 primesW4) w4
  primesES4 = 2:3:5:7:11:ts4 `sMinus` composites (drop 4 primesES4) s4
\end{verbatim}
}
\end{tcolorbox}
\end{minipage}
\end{center}
\vspace{-5mm}
\caption{Mounting Wheels on our sieves.}
\label{fig:wheel-Hamming}
\vspace{-1mm}
\end{figure}

Of course, wheels can be mounted on other programs, such as trial division and the Bird's Sieve of Fig.~\ref{fig:sieveBird}.  
As already observed in \cite{sieve09}, the trial division program does not gain so much from being equipped with the wheel $w_4$. 
The reason is that in such program, the wheel $w_4$ just prunes the stream of candidates primes to $S_4|^{11}$, 
but the erased numbers are those that trial division quickly recognises as non primes, as they are multiples 
of the first 4 primes.

In the Epilogue of \cite{sieve09}, the author hints to the fact that it is nontrivial to modify the Bird's sieve to 
support the wheel optimisation: 
this is true if we look at the elegant program of Fig.~\ref{fig:sieveBird} that makes use of 
list comprehension, but mounting a pre-computed wheel is almost trivial 
if we rewrite that program by following the same structure as all our 
sieves, based on explicit recursion. Adding suitable parameters to the function \code{composites},  
along the same lines of equipping our sieves with the wheel optimisation, 
we mount the $w_4$ wheel to the Bird's sieve as in Fig.~\ref{fig:sieveBird-w4}. 

\begin{figure}[b]
\begin{center}
\begin{minipage}{0.8\linewidth}
\begin{tcolorbox}[colback=white, boxrule=0pt, toprule=1pt, bottomrule=1pt, left=0pt]
{\small
\begin{verbatim}
   primes = 2:3:5:7:11:s4 `sMinus` composites (drop 4 primes) s4 where
      composites (p:ps) ss@(s:tss) = 
                   multiples p ss `unionP` composites ps tss
      multiples n ss = map (n*) (n:ss)\end{verbatim}
}
\end{tcolorbox}
\end{minipage}
\end{center}
\vspace{-6mm}
\caption{Bird's Sieve equipped with the wheel $w_4$.}
\label{fig:sieveBird-w4}
\vspace{-1mm}
\end{figure}

We observe that in this case, this optimisation is really significant, 
because also multiples of any prime $p$ are computed as $p\cdot S_4|^{p}$, rather than as $p\cdot \mathbb{N}_{\geq p}$, as 
in the original algorithm of Eratosthenes.

In other words, the program of Fig.~\ref{fig:sieveBird-w4} {\em is not}, strictly speaking, a 
{\em genuine} Sieve of Eratosthenes, but rather it implicitly encompasses in its computation the fourth iteration of the Euler's 
Sieve. 

The same holds for the O'Neill `faithful' Sieve of Eratosthenes, and this explains why this program 
is so efficient when equipped with the pre-computed wheel $w_4$, even though without $w_4$ it is clearly 
outperformed even by our stream based Euler's sieves, even without the $w_4$ optimisation (see Section~\ref{sec:expres} for 
details).

%% file: wheelONeal.tex
\section{Haskeller shall not Live by Streams Alone}
\label{sec:melissa}

As already discussed, the need of merging streams is the main bottleneck of our programs. 
As a matter of fact, extracting $n$ numbers in an ordered list from $m$ ordered lists is not linear in $n$ when $m$ is not constant, 
but rather ${\cal O}(n\cdot m)$. This problem looks impossible to be circumvented when generating composites as the 
union of a stream of ordered streams as we do in our programs.

The main virtue of the O'Neill Sieve~\cite{sieve09} 
is to circumvent this problem by using a priority queue to store/extract `efficiently' 
composites to be crossed off. In that program, composites are stored in a priority queue as pairs $(k, v)$, where the key $k$ is 
the next composite to be extracted from the iterator that generates all multiples of a certain prime, and the value $v$ is 
such iterator.

Both wheels (in the sieve {\sf W} of Fig.~\ref{fig:wheel-Sieve-2}) and 
the stream \code{es} generating the set $E_{k+1}$ 
(in the sieve {\sf ES} of Fig.~\ref{fig:efficient-euler-sieve})
are indeed iterators to generate composites, with the advantage, with respect to multiples, 
that two 
iterators generate disjoint streams of composites. 
Therefore, in the priority queue based O'Neill Sieve, 
we can easily replace iterators generating multiples 
with iterators generated by wheels or those generating sets $E_k$ 
leading to two Euler's sieve programs without the bottleneck of stream union.  

\begin{figure}[b]
\begin{center}
\begin{minipage}{0.80\linewidth}
\begin{tcolorbox}[colback=white, boxrule=0pt, toprule=1pt, bottomrule=1pt, left=0pt]
{\small
\begin{verbatim}
   sieve (c:cs) = c:sieve' cs ss (insertPQ (c*c) (tail es) emptyPQ) 
     where es  = map (c*) (c:cs)
           ss  = cs `sMinus` es
           sieve' cs@(c:tcs) ss tbl
              | c < n     = c : sieve' tcs  ss' tbl' 
              | otherwise = sieve' tcs ss tbl'' 
             where (n, m:ms) = minKeyValuePQ tbl
                   es    = map (c*) ss
                   ss'   = tail (ss `sMinus` es)  
                   tbl' = insertPQ (c*c) (tail es) tbl 
                   tbl'' = deleteMinAndInsertPQ m ms tbl
   primes = sieve [2..]
\end{verbatim}
}
\end{tcolorbox}
\end{minipage}
\end{center}
\vspace{-5mm}
\caption{{\sf EPQ}: the priority-queue based version of {\sf ES}.}
\label{fig:ESPQ}
\vspace{-1mm}
\end{figure}

In Fig.~\ref{fig:ESPQ}, we show the program that 
integrates the computation of the {\sf ES} sieve of Fig.~\ref{fig:efficient-euler-sieve}
into the O'Neill Sieve. We have just made the O'Neill code more compact 
and modified insertion into the table of composites: we do not just insert multiples of the tail of the stream to be sieved, 
but we insert a stream \code{es} that generates the set $E_{k+1}$. 
We have used variables \code{es} and \code{ss} with the same meaning as in the program 
of Fig.~\ref{fig:efficient-euler-sieve}, that is, 
at each invocation of the function \code{sieve'}, if the head \code{c} of the stream of prime candidates \code{cs}  
is the prime $p_k$, then \code{es} is $E_k$, and \code{ss} is $S_k|^{k+1}$. 
As before, $E_{k+1}$ is computed from $S_k$ that in turn is generated by the parameter \code{ss}.
%
%

Along the same lines, we can use wheels for the same purpose. The resulting program is in Fig.~\ref{fig:WPQ} 
(in this case we present the version equipped with the wheel $w4$). 
We incrementally compute wheels as in the program of Fig.~\ref{fig:wheel-Sieve-2} in such a way 
that at each invocation \code{sieve' cs@(c:tcs) w tbl}, if \code{c} is the prime $p_k$, then \code{w} 
is the wheel $w_k$.
Even though this program essentially consider the same sequences of composites as the one in Fig.~\ref{fig:ESPQ}, 
it wastes less memory thanks to the circular representation of wheels (see function \code{circ}). 
As we will see in Section~\ref{sec:expres}, this allows this program to compute efficiently the stream of primes for 
very large $n$, even though it is slightly less efficient than the program of Fig.~\ref{fig:ESPQ} for small values of $n$.

\begin{figure}[h]
\begin{center}
\begin{minipage}{0.80\linewidth}
\begin{tcolorbox}[colback=white, boxrule=0pt, toprule=1pt, bottomrule=1pt, left=0pt]
{\small
\begin{verbatim}
   sieve (c:cs) w = c:sieve' cs (nextWheel w c) 
                    (insertPQ (c*c) (circ (map (c*) w)) emptyPQ)  
     where sieve' cs@(c:tcs) w tbl
             | c < n     = c : sieve' tcs  w' tbl' 
             | otherwise = sieve' tcs w tbl'' 
            where (n, m:ms) = minKeyValuePQ tbl
                  w'    = nextWheel1 w c
                  tbl'  = insertPQ (c*c) (circ (map (c*) w)) tbl 
                  tbl'' = deleteMinAndInsertPQ (n+m) ms tbl 
   primes = 2:3:5:7:sieve s4 w4        
\end{verbatim}
}
\end{tcolorbox}
\end{minipage}
\end{center}
\vspace{-5mm}
\caption{{\sf WPQ}: the priority-queue based version of {\sf W}.}
\label{fig:WPQ}
\vspace{-1mm}
\end{figure}

%% file: expres.tex
\section{The Operation Was Successful, but the Patient Died}
\label{sec:expres}


In this section, we present an experimental evaluation of our programs, by comparing their running time to  
trial division~\cite{haskellPrimes}, the stream based Sieve of Eratosthenes by Richard Bird 
of Fig.~\ref{fig:sieveBird}, and the Melissa O'Neill faithful Sieve of Eratosthenes in \cite{sieve09}.   
The `unfaithful' Sieve of Eratosthenes in Fig.~\ref{fig:unfaithful-sieve} and the `na\"ive' Euler's Sieve in Fig.~\ref{fig:classical-sieve-of-euler} do not fit in our results as their running time are huge compared to those of above mentioned programs.
As an example, they compute $p_{2^{14}}$ in $\sim 2'$ and in $\sim 2'30''$ respectively (under the \code{GHCi} interpreter), and they reveal a (more than) quadratic experimental complexity (see also \cite{haskellPrimes}). As we report in our experiments, $p_{2^{16}}$ is computed in few seconds by all other programs we listed above (see Table~\ref{tab:interpreter}).

%
\subsection{Experimental Details}
Experimental results are reported in Tables~\ref{tab:interpreter}--\ref{tab:compilerPQ}. Tables~\ref{tab:interpreter} and  \ref{tab:compiler} report the running time of stream based programs, whereas those of priority-queue based programs are in 
Tables~\ref{tab:interpreterPQ} and~\ref{tab:compilerPQ}. 
Tables~\ref{tab:interpreter} and~\ref{tab:interpreterPQ} report the running time obtained running programs under the interpreter, 
whereas Tables~\ref{tab:compiler} and~\ref{tab:compilerPQ} report those of compiled programs.
  
In all tables, we call {\sf TD} the standard Trial Division algorithm~\cite{haskellPrimes}, {\sf BS} the Bird's Sieve in Fig.~\ref{fig:sieveBird}, 
and {\sf O'N} the faithful Sieve of Eratosthenes in \cite{sieve09}.
Our programs are referred to as in pictures, that is {\sf H} is the Hamming sieve of Fig.~\ref{fig:primes-hammings},  
{\sf W} is the wheel based sieve of Fig.~\ref{fig:wheel-Sieve-2}, {\sf ES} is the sieve of Fig.~\ref{fig:efficient-euler-sieve}, and {\sf EPQ} and {\sf WPQ} are sieves of Fig.~\ref{fig:ESPQ} and~\ref{fig:WPQ} based on a priority queue.
The superscript $^{4}$ denotes the program in which we mount the pre-computed wheel $w4$.

We show the running time computed by the \code{GHCi} interpreter by using the option \code{:set +s} and 
the running time of compiled programs using the \code{time} Unix command (\code{sys} plus \code{user} time).  
All experiments have been performed on an Intel i7 quad core, 2,5 GHz, 16Gb of RAM, under MacOSX 10.10.
In all tables, $n$ means `compute the prime $p_n$' and  
an asterisk $^*$ means that more than $10\%$ of the running time has been spent in system calls, that here means
essentially that the program has allocated a huge amount of memory and sometimes, the program has used virtual memory.

\begin{table*}[h]
\begin{center}
\begin{tabular}{c|rrr|rrr|rrr}
\toprule
\textbf{\textit{n}} &\textsf{TD} & \textsf{BS} & \textsf{BS}$^4$ & \textsf{H} & \textsf{W} & \textsf{ES} & \textsf{H$^{4}$} & \textsf{W$^{4}$}   & \textsf{ES}$^{4}$ \\ 
\midrule
  $2^{16}$ &    $6^8$  &    $12^3$ &    $2^5$ &    $2^9$ &    $2^5$ &    $2^4$ &     $1^5$ &    $1^4$ &  $1^1$  \\
  $2^{17}$ &   $18^5$  &    $34^6$ &    $6^0$ &    $7^2$ &    $6^0$ &    $5^8$ &     $3^6$ &    $3^9$ &  $3^0$ \\
  $2^{18}$ &   $47^7$  &  $1'31^3$ &   $19^5$ &   $15^5$ &   $14^1$ &   $13^5$ &     $8^5$ &   $10^1$ &  $7^1$ \\
  $2^{19}$ & $2'05^0$  &  $6'31^1$ &   $53^8$ &   $37^7$ &   $34^8$ &   $30^3$ &    $24^2$ &   $24^3$ & $19^8$ \\
  $2^{20}$ & $5'34^1$  & $11'08^6$ & $2'24^0$& $1'42^4$ & $1'29^8$ & $1'16^6$ &  $1'09^0$ & $1'05^7$ & $52^1$ \\
\bottomrule
\end{tabular}
\vspace{-1mm}
\caption{Running time: {\em stream programs} under the {\tt GHCi} interpreter and the \code{:set +s} option.}
\label{tab:interpreter}
\end{center}
\vspace{-4mm}
\end{table*}

\begin{table*}[b]
\begin{center}
\begin{tabular}{c|rrr|rrr|rrr}
\toprule
\textbf{\textit{n}} &\textsf{TD} & \textsf{BS} & \textsf{BS}$^4$ & \textsf{H} & \textsf{W} & \textsf{ES} & \textsf{H$^{4}$} & \textsf{W$^{4}$} & \textsf{ES}$^{4}$ \\
\midrule
  $2^{19}$ &     $8^9$ &    $10^9$ &     $3^0$ &     $12^1$ &    $2^7$ &     $4^2$ &      $6^3$ &    $2^1$ &      $1^7$\\
  $2^{20}$ &    $22^6$ &    $28^1$ &     $8^3$ &     $25^3$ &    $6^7$ &    $10^7$ &     $21^1$ &    $5^0$ &      $4^3$\\
  $2^{21}$ &    $58^7$ &  $1'22^2$ &    $25^2$ &   $1'21^4$ &   $17^0$ &    $24^5$ &     $49^0$ &   $14^1$ &     $11^8$\\
  $2^{22}$ &  $2'44^6$ &  $4'06^4$ &  $1'19^9$ & $^*9'27^5$ &   $46^5$ &  $1'00^4$ & $^*6'13^1$ &   $37^4$ &     $33^8$\\
  $2^{23}$ &  $7'03^5$ & $11'49^8$ &  $4'13^7$ &        $-$ & $2'16^8$ &$^*3'33^1$ &       $-$  & $1'50^8$ &   $1'36^4$\\
  $2^{24}$ & $19'13^5$ &       $-$ & $14'14^3$ &        $-$ & $6'33^9$ &       $-$ &       $-$  & $5'43^4$ & $^*5'51^8$\\
\bottomrule
\end{tabular}
\vspace{-1mm}
\caption{Running time: compiled {\em stream programs} using \code{time} Unix function (usr+sys).}
\label{tab:compiler}
\end{center}
\vspace{-4mm}
\end{table*}

\subsection{The Moral of our Experiments}
The Hamming Sieve {\sf H} is the slowest of our programs, but still much faster than the Bird's Sieve (even than {\sf BS}$^4$) 
and Trial Division when 
we run all these program under the \code{GHCi} interpreter. Strangely, {\sf H} gains much less than all other programs from 
compilation: in this case it is even slower than Trial Division and memory becomes quite early a big trouble for it.

Our stream based programs {\sf ES} and {\sf W} 
dramatically outperforms Trial Division, the Bird's Sieve, 
and even the Melissa O'Neill faithful Sieve of Eratosthenes without pre-computed wheels (even though 
{\sf O'N}
experimentally exhibits a better asymptotical complexity than {\sf ES$^4$} and {\sf W$^4$}). 
Remarkably, for small $n$, {\sf ES$^4$} is also faster than priority-queue based programs (for $n\leq 2^{18}$ running under 
the interpreter, and for $n\leq 2^{21}$ in the compiled arena).

Running time of {\sf W} is better than that of {\sf ES} when compiled, but once the pre-computed wheel $w4$ is mounted on 
both programs, {\sf ES$^4$} is slightly faster than {\sf W$^4$}, both in interpreted and compiled version. 
This probably depends on the overhead of (lazily) computing huge wheels, whose savings in terms of composites 
is not so important once the wheel $w_4$ has been mounted on. 
When memory becomes a critical resource, 
{\sf W} appears to be more parsimonious than {\sf ES} and {\sf W$^4$} can solve efficiently problem instances in which {\sf ES$^4$} 
severely slows down because of memory allocation.     

\begin{table*}[h]
\vspace{-3mm}
\begin{center}
\begin{tabular}{c|rrr|rrr}
\toprule
\textbf{\textit{n}} &\textsf{O'N} & \textsf{WPQ} & \textsf{EPQ} & \textsf{O'N$^{4}$} & \textsf{WPQ$^{4}$} & \textsf{EPQ$^{4}$}\\ 
\midrule
  $2^{18}$ &     $58^5$ &   $19^3$ &   $20^7$ &   $11^1$ &    $8^6$ &    $8^0$ \\
  $2^{19}$ &   $2'08^9$ &   $41^8$ &   $44^8$ &   $24^8$ &   $18^9$ &   $17^6$ \\
  $2^{20}$ &   $4'21^5$ & $1'29^4$ & $1'35^1$ &   $52^5$ &   $39^7$ &   $38^5$ \\
  $2^{21}$ &   $9'08^6$ & $2'45^8$ & $3'01^6$ & $1'55^4$ & $1'27^8$ & $1'24^6$ \\
  $2^{22}$ &      $-$   & $5'55^7$ & $7'04^7$ & $4'14^4$ & $2'57^9$ & $2'53^3$ \\
\bottomrule
\end{tabular}
\vspace{-2mm}
\caption{Running time: {\em Priority queue programs} under the {\tt GHCi} interpreter.}
\label{tab:interpreterPQ}
\end{center}
\vspace{-4mm}
\end{table*}
  
Nevertheless, experimental results are a bit disappointing from our point of view. 
Our best programs, {\sf EPQ$^4$} and {\sf WPQ$^4$},  
are definitively the fastest when we execute all programs inside the \code{GHCi} interpreter (Table~\ref{tab:interpreterPQ}), 
but they fail to be convincingly faster than {\sf O'N$^4$}, when programs are compiled  (Table~\ref{tab:compilerPQ}).  
Similar to {\sf ES}, {\sf EPQ$^4$} severely slows down computing primes beyond $p_{2^{24}}$ because of memory allocation.
By contrast, {\sf WPQ} and {\sf WPQ$^4$} are still quite efficient in the computation of the first $2^{25}$ primes, 
when also {\sf O'N$^4$} goes trough a significant performance degradation (Table 4).
{\sf WPQ$^4$} succeeds in computing primes beyond $2^{25}$, when all other programs are killed by the operating 
system, due to excessive memory requirements.

As we observed in Section~\ref{sec:pitStop}, {\sf O'N$^4$} encompasses the fourth iteration of the Euler's 
Sieve, and this explains the huge speed-up with respect to {\sf O'N}.
By contrast, for our {\sf WPQ} and {\sf EPQ}, mounting the wheel $w_4$ means just saving their first 4 recursive calls, 
and this is not as important for them as it is for {\sf O'N}. In particular, for wheel based sieves ({\sf W} and {\sf WPQ}) 
mounting the wheel $w_4$ just prunes the stream of candidate primes to be examined, 
as they efficiently compute the first 4 small wheels.
Indeed, {\sf WPQ} is slower but competitive with all prime generators mounting the wheel $w4$ (see Table~\ref{tab:compilerPQ}) and we 
can consider {\sf WPQ$^4$} our {fast prime generator} as it is fast enough for small $n$ and the most resistant to performance degradation for large $n$.

\begin{table*}[h]
\begin{center}
\begin{tabular}{c|rrr|rrr}
\toprule
\textbf{\textit{n}} &\textsf{O'N} & \textsf{WPQ} & \textsf{EPQ} & \textsf{O'N$^{4}$} & \textsf{WPQ$^{4}$} & \textsf{EPQ$^{4}$}\\ 
\midrule
$2^{21}$       &   $34^8$ &   $15^0$  &     $23^1$ &     $11^8$ &     $12^3$ &      $13^2$\\
$2^{22}$       & $1'11^5$ &   $31^8$  &     $51^1$ &     $25^2$ &     $27^3$ &      $28^2$\\
$2^{23}$       &$^*2'49^6$& $1'15^1$  & $^*2'25^4$ &     $56^3$ &   $1'03^3$ &    $1'00^7$\\
$3\cdot 2^{22}$&$^*5'34^0$& $1'57^8$  & $^*5'15^1$ &   $1'29^9$ &   $1'38^5$ &    $1'31^6$\\
$2^{24}$       &        - & $2'33^3$  &          - &   $2'11^3$ &   $2'10^0$ &    $2'08^2$\\
$7\cdot 2^{22}$&        - & $5'55^9$  &          - & $^*5'43^7$ &   $5'27^9$ & $^*10'22^9$\\
$2^{25}$       &        - &$^*8'45^9$ &          - & $^*8'21^7$ & $^*7'34^2$ &          - \\
\bottomrule
\end{tabular}
\vspace{-2mm}
\caption{Running time: compiled {\em priority queue programs}.}
\label{tab:compilerPQ}
\end{center}
\vspace{-7mm}
\end{table*}

%% file: conclu.tex
\section{Conclusion and Future Work}
\label{sec:conclu}

We have presented three stream based Haskell implementation of the Euler's Sieve.
The resulting programs are pretty efficient with respect to other stream based prime sieves.
Their faithfulness with respect to the original algorithm is up to the overhead of union operation 
over streams: even though we succeed in generating each composite to be crossed off just once, a composite number 
will be compared several times before joining the list of composites.
To overcome such problem, we have integrated ideas behind two of our Euler's Sieves in the 
priority queue structure of the O'Neill's `faithful' Sieve of Eratosthenes. 
Our sieve of Fig.~\ref{fig:WPQ} based on wheels and a priority queue is our fast prime generator as it results 
both fast enough on small instances and the most robust to performance degradation because of memory allocation on large  
instances. 

Several interesting questions still remains open. 
First of all, it would be interesting to investigate why the Hamming Sieve is so `resistant'  
to compiler optimisations. 
Moreover, it would be interesting to look for some advanced data structure 
that could improve its performances, since it appears not trivial use priority queues to speed-up 
its computation. 

An intriguing question is about using wheels to cross off composites in an array based imperative 
implementation of the Euler's Sieve. Of course, the lazy computation of wheels 
is crucial as already observed by~\cite{runciman97}, 
and this could be hard to code properly in an eager imperative language, but the advantage would be to avoid 
additional data structures such as a double linked list in \cite{js90}. 

Finally, it would be interesting to carefully look for optimisations to make our faster sieves 
competitive with the current prime generator in the \code{Data.Number} Haskell library.
%